\documentclass[11pt]{article}
\usepackage{graphicx}
\usepackage{amsfonts}
\usepackage{theorem}
\newtheorem{Lem}{Lemma}[section]

\newtheorem{The}[Lem]{Theorem}
\newtheorem{Prop}[Lem]{Proposition}
\newtheorem{Cor}[Lem]{Corollary}

\newtheorem{Rem}[Lem]{Remark}

\newcommand{\qed}{\hbox{\rule{6pt}{6pt}}}

\begin{document}
\title{Precise estimates of bounds on relative operator entropies}
\author{Shigeru Furuichi\footnote{E-mail:furuichi@chs.nihon-u.ac.jp}\\
{\small Department of Information Science,}\\
{\small College of Humanities and Sciences, Nihon University,}\\
{\small 3-25-40, Sakurajyousui, Setagaya-ku, Tokyo, 156-8550, Japan}}
\date{}
\maketitle
{\bf Abstract.} Recently, Zou obtained the generalized results on the bounds for Tsallis relative operator entropy. In this short paper, we give precise bounds for Tsallis relative operator entropy. We also give precise bounds of relative operator entropy.

\vspace{3mm}


\vspace{3mm}
\vspace{3mm}

\section{Introduction}
As one of extensions for relative operator entropy 
$$
S(A|B) \equiv A^{1/2}\log \left(A^{-1/2}BA^{-1/2}\right) A^{1/2}
$$
introduced by Fujii and Kamei \cite{FK} (similar quantity was introduced by Belavkin and Staszewski in \cite{BS}), for invertible positive operators $A$ and $B$,
Tsallis relative operator entropy was defined as \cite{YKF}:
$$
T_r(A|B) \equiv A^{1/2}\ln_r \left(A^{-1/2}BA^{-1/2}\right) A^{1/2} = \frac{A\sharp_r B -A}{r}
$$
for $r \in (0,1]$ and  invertible positive operators $A$ and $B$, where one-parameter extended logarithmic function $\ln_r (\cdot)$ is defined as $\ln_r x \equiv \frac{x^r-1}{r}$ which uniformly converges to the usual logarithmic function $\log x$ when $r \to 0$. Here the weighted geometric mean is defined by $A\sharp_r B \equiv A^{1/2}\left(A^{-1/2}BA^{-1/2}\right)^r A^{1/2}$ for  $r \in [0,1]$ and  invertible positive operators $A$ and $B$. 
Tsallis relative operator entropy $T_r(A|B) $ is a one-parameter extension of  relative operator entropy $S(A|B)$ in the sense that 
$$
\lim_{r\to 0} T_r(A|B) = S(A|B).
$$

Recently, Zou obtained new operator inequalities
on Tsallis relative operator entropy $T_{r}(A|B)$ in \cite{Zou}.
His results generalized our previous results \cite{FYK}.

\begin{The}  {\bf (\cite{Zou})} \label{Zou_theorem}
For $a >0$, $r \in (0,1]$, $t \in [0,1]$ and two invertible positive operators $A$ and $B$, the following inequalities hold:
\begin{equation} \label{Zou_ineq}
c_3 A\sharp_r B -c_1A\natural_{r-1}B-c_2A \leq T_r(A|B) \leq c_1 B+c_2A\sharp_r B-c_3 A
\end{equation}
where $A \natural_r B \equiv A^{1/2} \left( A^{-1/2} BA^{-1/2}\right)^rA^{1/2}$ is defined for $r \in \mathbb{R}$ and invertible positive operators $A$ and $B$.
For simplicity, we also set $c_i \equiv c_i(a,r,t),\,\, (i=1,2,3)$ where
$$
c_1(a,r,t) \equiv \frac{r a^{r-1}}{d(a,r,t)},\,\,
c_2(a,r,t) \equiv \frac{t(a^r -1)}{d(a,r,t)},\,\,
c_3(a,r,t) \equiv \frac{r a^r +(t-1)(a^r -1)}{d(a,r,t)},
$$
with $d(a,r,t) \equiv r \left\{ ta^r +(1-t)\right\}$.
\end{The}

As remarked in \cite{Zou}, the inequalities (\ref{Zou_ineq}) recovers
the inequalities \cite{FYK}:
\begin{equation} \label{FYK2005_ineq}
A\sharp_r B -\frac{1}{a}A \natural _{r-1}B + \left(\ln_r \frac{1}{a}\right) A \leq T_r(A|B) \leq
\frac{1}{a} B - \left(\ln_r \frac{1}{a}\right) A\sharp_r B -A
\end{equation}
putting $t=1$ in the inequalities (\ref{Zou_ineq}), since $c_1 =\frac{1}{a}$, $c_2=-\ln_r \frac{1}{a}$ and $c_3 =1$. That is, the inequalities (\ref{Zou_ineq}) generalized the inequalities (\ref{FYK2005_ineq}).
Again taking the limit as $r \to 0$ in the inequalities (\ref{FYK2005_ineq}),
we recover the following inequalities shown by Furuta in \cite{Furuta}:
\begin{equation} \label{Furuta1993_ineq}
(1-\log a) A -\frac{1}{a} AB^{-1}A \leq S(A|B) \leq (\log a -1) A +\frac{1}{a} B.
\end{equation}
In this short paper, we give the further precise inequalities concerning the inequalities
(\ref{Zou_ineq}) and  (\ref{FYK2005_ineq}).


\section{Main results}
\begin{The} \label{main_theorem}
Let $A$ and $B$ be positive invertible operators on a Hilbert space, and $a > 0$, $r \in (0,1]$, $t \in [0,1]$.  Then the following inequalities (i) and (ii) hold.
\begin{itemize}
\item[(i)] If $0 < a \leq 1$, then 
\begin{eqnarray*}
A\sharp_r B -\frac{1}{a}A \natural _{r-1}B + \left(\ln_r \frac{1}{a}\right) A &\leq&
 c_3 A\sharp_r B -c_1A\natural_{r-1}B-c_2A \\
&\leq& T_r(A|B)  \\
&\leq& c_1 B+c_2A\sharp_r B-c_3 A \\
&\leq& \frac{1}{a} B -\left(\ln_r \frac{1}{a}\right) A\sharp_r B -A
\end{eqnarray*}
\item[(ii)] If $a \geq 1$, then 
\begin{eqnarray*}
 c_3 A\sharp_r B -c_1A\natural_{r-1}B-c_2A &\leq&
A\sharp_r B -\frac{1}{a}A \natural _{r-1}B +\left(\ln_r \frac{1}{a}\right)A \\
&\leq& T_r(A|B)  \\
&\leq&  \frac{1}{a} B -\left(\ln_r \frac{1}{a}\right) A\sharp_r B -A \\
&\leq& c_1 B+c_2A\sharp_r B-c_3 A
\end{eqnarray*}
\end{itemize}
\end{The}
{\it Proof:}
Thanks to the theory by Kubo and Ando \cite{KA}, if we use the notion of the representing function $f_m(x)=1mx$ for operator mean $m$, then 
the scalar inequality $f_m(x) \leq f_n(x)$ (for $x >0 $) is equivalent to
the operator inequality $AmB \leq AnB$ for all positive operators $A$ and $B$.
Therefore we have only to consider the scalar inequalities to prove the operator inequalities in this theorem.
In addition, the second inequality and the third inequality in both (i) and (ii) have already proven in the inequalities (\ref{Zou_ineq}) and (\ref{FYK2005_ineq}). We thus prove the first inequality and the last inequality in both (i) and (ii) .
To do so, we firstly set the function $l(a,r,t,x)$ as
$$
l(a,r,t,x) \equiv c_3(a,r,t) x^r -c_1(a,r,t) x^{r-1}-c_2(a,r,t). 
$$
By elementary calculations, we have
$$
\frac{dl(a,r,t,x)}{dt} = \frac{(a^r -1)}{r\left\{ t a^r+(1-t)\right\}^2}h(a,r,x),
$$
where
$
h(a,r,x) = (1-r) a^r x^r + r a^{r-1}x^{r-1} -1.
$
Since we have $\frac{dh(a,r,x) }{dx} = r(1-r)a^{r-1}x^{r-2}(ax-1)$, we have $h(a,r,x) \geq h(a,r,1/a) =0$ for any $a >0$, $x>0$ and $r \in (0,1]$.
Thus we have $\frac{dl(a,r,t,x)}{dt} \leq 0$ if $0 < a \leq 1$, and $\frac{dl(a,r,t,x)}{dt} \geq 0$ if $a \geq 1$. Therefore we have $l(a,r,t,x) \geq l(a,r,1,x) $ if $0 < a \leq 1$, and $l(a,r,t,x) \leq l(a,r,1,x) $ if $a \geq 1$.
When $t=1$, it follows that $c_1=\frac{1}{a}$, $c_2= -\ln_r \left( \frac{1}{a} \right)$ and $c_3 = 1$. Thus the first inequalities in both (i) and (ii) have been proven.

We also set the function $u(a,r,t,x)$ as
$$
u(a,r,t,x) \equiv c_1(a,r,t)x +c_2(a,r,t)x^r -c_3(a,r,t).
$$
By elementary calculations, we have
$$
\frac{du(a,r,t,x)}{dt} = \frac{(a^r -1)}{r\left\{ t a^r+(1-t)\right\}^2}g(a,r,x),
$$
where
$
g(a,r,x) =  x^r  -r a^{r-1}x -(1-r)a^r.
$
Since $\frac{dg(a,r,x) }{dx} =r(x^{r-1}-a^{r-1})$, we have $g(a,r,x) \leq g(a,r,a)=0$
for any $a >0$, $x>0$ and $r \in (0,1]$. Thus we have $\frac{du(a,r,t,x)}{dt} \geq 0$ if $a \leq 1$, and $\frac{du(a,r,t,x)}{dt} \leq 0$ if $a \geq 1$. Therefore we have
$u(a,r,t,x) \leq u(a,r,1,x) $ if $a \leq 1$, and $u(a,r,t,x) \geq u(a,r,1,x) $ if $a \geq 1$.
Thus the last inequalities in both (i) and (ii) have been proven.

\hfill \qed

\begin{Rem}
Theorem \ref{main_theorem} shows that
the inequalities (\ref{Zou_ineq}) give tight bounds of $T_{r}(A|B)$ when $0 < a \leq 1$, however the inequalities (\ref{FYK2005_ineq}) give tight bounds of $T_{r}(A|B)$ when $a \geq 1$.
\end{Rem}

In the paper \cite{Zou}, Zou obtained the lower bound of $T_{-r}(A|B)$ for $r \in (0,1]$ as
\begin{equation} \label{Zou_ineq_lower00}
(a^r- \ln_r a) A -a^{r-1} AB^{-1} A \leq T_{-r}(A|B).
\end{equation}
We also obtain the different upper bound of Tsallis relative operator entropy as:
\begin{equation} \label{Zou_upper_lower00}
T_{r}(A|B) \leq (\ln_r a -a^r) A + a^{r-1} B.
\end{equation}
The above inequality (\ref{Zou_upper_lower00}) can be proven by the scalar inequality
\begin{equation} \label{Zou_upper_lower01}
\ln_r x \leq \ln_r a -a^r +a^{r-1} x
\end{equation}
which is obtained by putting $z = \frac{x}{a}$ in the fundamental inequality $\ln_r z \leq z-1$ for $z >0$ and $r \in (0,1].$
Thus we have the following proposition, taking account for the inequalities (\ref{Zou_ineq_lower00}) and (\ref{Zou_upper_lower00}) with the following inequalities \cite{Zou,FYK}:
\begin{equation} \label{FYK_TST_ineq}
T_{-r}(A|B) \leq S(A|B) \leq T_{r}(A|B).
\end{equation}
\begin{Prop}
For $a >0$, $r \in (0,1]$ and invertible positive operators $A$ and $B$, 
the following inequalities hold.
\begin{eqnarray} \label{prop1_ineq}
 (a^r- \ln_r a) A -a^{r-1} AB^{-1} A &\leq& T_{-r}(A|B)
\leq S(A|B) \nonumber \\
&\leq& T_{r}(A|B) \leq (\ln_r a -a^r) A + a^{r-1} B.
\end{eqnarray}
\end{Prop}

In the paper \cite{FYK}, we also obtained the following inequality
\begin{equation} \label{FYK_upper_lower00}
T_r(A|B) \leq \frac{1}{a} B -A - \left(\ln_r \frac{1}{a}\right) A \sharp_r B
\end{equation}
from the scalar inequality $\ln_r x \leq \frac{x}{a} -1 - \left(\ln_r \frac{1}{a}\right) x^r$.
By the slight modification of this scalar inequality, we can obtain the following inequality:
\begin{equation} \label{FYK_lower_lower00}
T_{-r}(A|B) \geq A -\frac{1}{a} AB^{-1} A +\left(\ln_r \frac{1}{a}\right)A\natural_{-r} B.
\end{equation}
Thus we also have the following proposition.
\begin{Prop}
For $a >0$, $r \in (0,1]$ and invertible positive operators $A$ and $B$, 
we have
\begin{eqnarray} \label{prop2_ineq}
A -\frac{1}{a} AB^{-1} A +\left(\ln_r \frac{1}{a}\right) A\natural_{-r} B &\leq& T_{-r}(A|B)
\leq S(A|B) \nonumber \\
&\leq& T_{r}(A|B) \leq \frac{1}{a} B -A - \left(\ln_r \frac{1}{a}\right) A \sharp_r B.
\end{eqnarray}
\end{Prop}

One may have an interest in the precise ordering on two inequalities (\ref{prop1_ineq}) and (\ref{prop2_ineq}). Then we can show the following corollary by Theorem \ref{main_theorem}.

\begin{Cor} \label{cor01}
For $a >0$, $r \in (0,1]$ and invertible positive operators $A$ and $B$, 
we have the following inequalities.
\begin{itemize}
\item[(i)] If $0 < a \leq 1$, then we have
\begin{eqnarray}
A -\frac{1}{a} AB^{-1} A +\left(\ln_r \frac{1}{a}\right) A\natural_{-r} B &\leq&  (a^r- \ln_r a) A -a^{r-1} AB^{-1} A \nonumber \\
&\leq &T_{-r}(A|B) \leq S(A|B) \leq T_{r}(A|B)  \nonumber \\
&\leq&   (\ln_r a -a^r) A + a^{r-1} B \leq \frac{1}{a} B -A - \left(\ln_r \frac{1}{a}\right) A \sharp_r B.\nonumber
\end{eqnarray}
\item[(ii)]If $a \geq 1$, then we have
\begin{eqnarray}
(a^r- \ln_r a) A -a^{r-1} AB^{-1} A &\leq& A -\frac{1}{a} AB^{-1} A +\left(\ln_r \frac{1}{a}\right) A\natural_{-r} B  \nonumber \\
&\leq &T_{-r}(A|B) \leq S(A|B) \leq T_{r}(A|B)  \nonumber \\
&\leq& \frac{1}{a} B -A -\left(\ln_r \frac{1}{a}\right) A \sharp_r B  \leq  (\ln_r a -a^r) A + a^{r-1} B.\nonumber
\end{eqnarray}
\end{itemize}
\end{Cor}
{\it Proof:}
When $a \leq 1$, we put $t=0$ in Theorem \ref{main_theorem}. 
Then we have $c_1=a^{r-1}$, $c_2=0$, $c_3=a^r -\ln_r a$ and we have the last inequality in (i):
$$
 a^{r-1} B + (\ln_r a -a^r) A  \leq \frac{1}{a} B -A - \left(\ln_r \frac{1}{a}\right) A\sharp_r B
$$
which is equivalent to
$$
a^{r-1} x +(\ln_r a -a^r)  \leq \frac{x}{a} - \left(\ln_r \frac{1}{a}\right) x^r -1.
$$
From this inequality, we have
$$
-a^{r-1} \frac{1}{x} -(\ln_r a -a^r) \geq 1-\frac{1}{ax}+\left(\ln_r \frac{1}{a}\right) x^{-r}
$$
which is equivalent to
$$
A -\frac{1}{a} AB^{-1} A +\left(\ln_r \frac{1}{a}\right) A\natural_{-r} B \leq  (a^r- \ln_r a) A -a^{r-1} AB^{-1} A.
$$
Thus (i) was proven. (ii) can be proven by similar way.

\hfill \qed

If we take $a=1$ in Corollary \ref{cor01}, then we recover the inequalities:
$$
A-AB^{-1}A \leq T_{-r}(A|B) \leq S(A|B) \leq T_{r}(A|B) \leq B-A
$$
which were given in the paper \cite{Zou}.
If we also take the limit $r \to 0$ in Corollary \ref{cor01}, then we recover the inequalities (\ref{Furuta1993_ineq}).


We can show another bounds of the relative operator entropy $S(A|B)$ under the restricted conditions. For this purpose, we prepare the following lemma.
\begin{Lem} \label{lemmaforfinaltheorem}
\begin{itemize}
\item[(i)] For $r \in [1/2, 1]$, $\alpha \in [-1,0) \cup (0,1]$ and $x \geq 1$, we have
$$
0 \leq 1- \frac{1}{x} \leq \frac{x^{-r} -1}{-r} \leq \frac{2(x-1)}{x+1} \leq  
\frac{2(x^{\alpha} -1)}{\alpha (x^{\alpha} +1)} \leq \log x \leq \frac{x^{\alpha} -1}{\alpha x^{\alpha /2}} \leq \frac{x-1}{\sqrt{x}} \leq \frac{x^r -1}{r} \leq x-1.
$$
\item[(ii)] For   $r \in [1/2, 1]$, $\alpha \in [-1,0) \cup (0,1]$ and $0 < x \leq 1$, we have
$$
1-\frac{1}{x} \leq \frac{x^{-r} -1}{-r} \leq \frac{x-1}{\sqrt{x}} \leq \frac{x^{\alpha} -1}{\alpha x^{\alpha /2}} \leq  \log x \leq  
\frac{2(x^{\alpha} -1)}{\alpha (x^{\alpha} +1)}  \leq \frac{2(x-1)}{x+1} \leq \frac{x^r -1}{r} \leq x-1 \leq 0.
$$
\end{itemize}
\end{Lem}

{\it Proof:}
The first inequality in (i) is trivial. 
We put $f(r,x) \equiv  \frac{x^r -1}{r} - \frac{x-1}{\sqrt{x}}$. Then we have
$\frac{df(r,x)}{dx} = \frac{2x^{r+1/2}-x-1}{2x^{3/2}} \geq 0$ for $r \geq 1/2$ and $x \geq 1$. Thus we have $f(r,x) \geq f(r,1) =0$ which proves the eighth inequality in (i). Since we have $\frac{d}{dr} \left( \frac{x^{-r} -1}{-r} \right) 
= \frac{1-x^r + r\log x}{x^r r^2} \leq 0$ (which implies the second inequality), we have $\frac{x^{-r} -1}{-r} \leq \frac{x^{-1/2} -1}{-1/2}$ if $r \geq 1/2$. When $x \geq 1$, we can prove the inequality
$\frac{2(x-1)}{x+1} \geq \frac{x^{-1/2} -1}{-1/2}$ with elementary calculations. Thus we have the third inequality of (i). 
We note that we have the inequalities
\begin{equation}\label{lemmaforfinaltheorem_ineq_standard00}
 \frac{2(x-1)}{x+1} \leq \log x \leq  \frac{x-1}{\sqrt{x}} ,\quad (x \geq 1).
\end{equation}
We put $g(\alpha,x) \equiv \frac{2(x^{\alpha} -1)}{\alpha (x^{\alpha} +1)}$ for $x \geq 1$. Since we have $g(-\alpha,x) = g(\alpha,x)$, we consider the case $0 < \alpha \leq 1$.
Then we have $\frac{d g(\alpha,x)}{d\alpha} = \frac{2\left\{ x^{\alpha} \log x^{2\alpha} -\left(x^{2\alpha}-1\right)\right\}}{\alpha^2 (x^{\alpha} +1)^2} \leq 0$, by replacing $x$ in the second inequality of (\ref{lemmaforfinaltheorem_ineq_standard00}) with $x^{2\alpha}$. Thus we have
$g(1,x) \leq g(\alpha,x) \leq \lim_{\alpha \to 0} g(\alpha,x)=\log x$ which imply the fourth and fifth inequalities of (i). We also put $h(\alpha,x) \equiv \frac{x^{\alpha} -1}{\alpha x^{\alpha /2}}$  for $x \geq 1$.  Since we have $h(-\alpha,x) = h(\alpha,x)$, we consider the case $0 < \alpha \leq 1$. Then we have $\frac{d h(\alpha,x)}{d\alpha} = \frac{\left( x^{\alpha} +1\right)\log x^{\alpha} -2\left(x^{\alpha} -1 \right) }{2\alpha^{2} x^{\alpha /2}} \geq 0$, by replacing $x$ in the first inequality of (\ref{lemmaforfinaltheorem_ineq_standard00}) with $x^{\alpha}$.  Thus we have $\log x = \lim_{\alpha \to 0} h(\alpha,x) \leq h(\alpha,x) \leq h(1,x)$ which imply the sixth and seventh inequalities of (i).
The last inequality of (i) comes from $\frac{d}{dr} \left( \frac{x^r -1}{r}\right)  = \frac{x^r}{r^2}\left(x^{-r} -1 +r \log x \right) \geq 0$.
Replacing $x$ in (i) with $1/x$, then we have the inequalities (ii) with elementary calculations.

\hfill \qed

\begin{Rem}
If $0 < r <1/2$, then we have
$$\lim_{x \to \infty} \frac{\frac{x^r-1}{r}}{\frac{x-1}{\sqrt{x}}} = \lim_{x \to \infty}
\frac{x^{r+1/2}-x^{1/2}}{r(x-1)} = \lim_{x\to \infty} \frac{(r+1/2)x^{r-1/2}-1/2x^{-1/2}}{r} =0.$$
This means that there exists $x>0$ such that $\frac{x^r-1}{r} \leq \frac{x-1}{\sqrt{x}}$ when $0< r < 1/2$. Actually, if we take $r=10/21$, then we have $\frac{x^r-1}{r} - \frac{x-1}{\sqrt{x}} = -23829.6$  when $x= 4 \times 10^{13}.$
For the same case $r=10/21$, we have $\frac{x^r-1}{r} - \frac{x-1}{\sqrt{x}} = 16866.3$ when $x= 3 \times 10^{13}.$ 
We should consider the case that $r$ is small. If we take $r=0.001$, then we have
 $\frac{x^r-1}{r} - \frac{x-1}{\sqrt{x}} = 7.26842\times 10^{-9}$ when $x= 1.005$, and   $\frac{x^r-1}{r} - \frac{x-1}{\sqrt{x}} = -2.66798\times 10^{-8}$ when $x= 1.015$.  Therefore  we may conclude that there is no ordering between $\frac{x^r-1}{r}$ and $\frac{x-1}{\sqrt{x}}$ for $0< r < 1/2$ and $x > 1$.
We also do not have the ordering between $\frac{x^{-r}-1}{-r}$ and $\frac{x-1}{\sqrt{x}}$ for $0< r < 1/2$ and $0< x < 1$.

If $0< r < 1/2$, then we also have
$$
\lim_{x \to \infty}  \frac{\frac{2(x-1)}{x+1}}{\frac{x^{-r}-1}{-r}}=  \lim_{x \to \infty} \frac{-2r(x-1)}{(x+1)(x^{-r}-1)} =\lim_{x \to \infty} \frac{2r}{1-x^{-r}+r (x^{-r-1} + x^{-r})}= 2r <1.
$$
This also means that there exists $x>0$ such that $\frac{2(x-1)}{x+1} \leq \frac{x^{-r}-1}{-r}$ when $0< r < 1/2$. Actually, if we take $r=10/21$, then we have $\frac{2(x-1)}{x+1} - \frac{x^{-r}-1}{-r} = -0.00681889$ when $x= 6 \times 10^2$. For the same case $r=10/21$, we have $\frac{2(x-1)}{x+1} - \frac{x^{-r}-1}{-r} = 0.000907845$ when $x=5 \times 10^2$. We also consider the case that $r$ is small.  If we take $r=0.001$, then we have $\frac{2(x-1)}{x+1} - \frac{x^{-r}-1}{-r} = 2.09877\times 10^{-9}$ when $x= 1.005$, and $\frac{2(x-1)}{x+1} - \frac{x^{-r}-1}{-r} = -1.6419 \times 10^{-7}$ when $x= 1.015$.
Therefore we may conclude that there is no ordering between $\frac{2(x-1)}{x+1}$ and $\frac{x^{-r}-1}{-r} $ for $0< r < 1/2$ and $x > 1$.
We also do not have the ordering between $\frac{2(x-1)}{x+1}$ and  $\frac{x^{r}-1}{r}$ for $0< r < 1/2$ and $0< x < 1$.
\end{Rem}

\begin{The} \label{final_theorem}
\begin{itemize}
\item[(i)] For $r \in [1/2, 1]$, $\alpha \in [-1,0) \cup (0,1]$ and $0 < A \leq B$, we have
\begin{eqnarray*}
0 &\leq& T_{-1}(A|B) \leq T_{-r}(A|B) \leq 
2A^{1/2}\left\{I -A^{1/2}\left(\frac{A+B}{2}\right)^{-1}A^{1/2} \right\}A^{1/2}  \\
&\leq& \frac{2}{\alpha}A^{1/2}\left\{ I-A^{1/2}\left( \frac{A+A\natural_{\alpha}B}{2}\right)^{-1}A^{1/2}\right\} A^{1/2}
\leq S(A|B) \\
&\leq& \frac{A\natural_{\alpha/2}B -A\natural_{-\alpha/2}B}{\alpha}
\leq  A\sharp_{1/2}B -A\left(A^{-1}\sharp_{1/2}B^{-1} \right)A \leq T_r(A|B) \leq T_1(A|B).
\end{eqnarray*}
\item[(ii)] For $r \in [1/2, 1]$, $\alpha \in [-1,0) \cup (0,1]$ and $0 < B \leq A$, we have
\begin{eqnarray*}
T_{-1}(A|B) &\leq& T_{-r}(A|B) \leq A\sharp_{1/2}B -A\left(A^{-1}\sharp_{1/2}B^{-1} \right)A \leq \frac{A\natural_{\alpha/2}B -A\natural_{-\alpha/2}B}{\alpha} \\
&\leq& S(A|B) 
\leq \frac{2}{\alpha}A^{1/2}\left\{ I-A^{1/2}\left( \frac{A+A\natural_{\alpha}B}{2}\right)^{-1}A^{1/2}\right\} A^{1/2} \\
&\leq&  
 2A^{1/2}\left\{I -A^{1/2}\left(\frac{A+B}{2}\right)^{-1}A^{1/2} \right\}A^{1/2}
 \leq T_{r}(A|B) \leq T_{1}(A|B) \leq 0.
\end{eqnarray*}
\end{itemize}
\end{The}

{\it Proof:}
We note that $T_1(A|B) = B-A$ and $T_{-1}(A|B) = A-AB^{-1}A$.
We apply (i) of Lemma \ref{lemmaforfinaltheorem}. Then the condition $x \geq 1$ implies $A^{-1/2}BA^{-1/2} \geq I$, and hence (i) of the present theorem requires the conditions $0 < A \leq B$. By some calculations, we obtain this theorem thanks to the theory of operator mean by Kubo and Ando \cite{KA}. 
(ii) of the present theorem can be proven similarly by applying (ii) of Lemma \ref{lemmaforfinaltheorem}. 
 
\hfill \qed

If we relax the condition $1/2 \leq r \leq 1$, then we have the following result.
\begin{Cor}  \label{final_corollary}
\begin{itemize}
\item[(i)] For $r \in  (0,1]$ and $0 < A \leq B$, we have
\begin{eqnarray*}
0 &\leq& T_{-1}(A|B) \leq T_{-r}(A|B) \leq
 \frac{2}{r}A^{1/2}\left\{ I-A^{1/2}\left( \frac{A+A\natural_{r}B}{2}\right)^{-1}A^{1/2}\right\} A^{1/2}\\
&\leq& S(A|B) 
\leq \frac{A\natural_{r/2}B -A\natural_{-r/2}B}{r}
\leq  T_r(A|B) \leq T_1(A|B).
\end{eqnarray*}
\item[(ii)] For $r \in  (0,1]$ and $0 < B \leq A$, we have
\begin{eqnarray*}
T_{-1}(A|B) &\leq& T_{-r}(A|B)  \leq \frac{A\natural_{r/2}B -A\natural_{-r/2}B}{r} 
\leq S(A|B)\\ 
&\leq& \frac{2}{r}A^{1/2}\left\{ I-A^{1/2}\left( \frac{A+A\natural_{r}B}{2}\right)^{-1}A^{1/2}\right\} A^{1/2} 
 \leq T_{r}(A|B) \leq T_1(A|B) \leq 0.
\end{eqnarray*}
\end{itemize}
\end{Cor}

{\it Proof:}
The proof can be done by using Theorem \ref{final_theorem} and the following scalar inequalities.
\begin{itemize}
\item[(i)] For $r \in (0,1]$ and $x \geq 1$, we have
$$
\frac{x^{-r} -1}{-r} \leq  \frac{2(x^{r} -1)}{r (x^{r} +1)} \leq \log x \leq \frac{x^{r} -1}{r x^{r /2}}  \leq \frac{x^r -1}{r}.
$$ 
\item[(ii)] For $r \in (0,1]$ and $0 < x \leq 1$, we have
$$
\frac{x^{-r} -1}{-r} \leq \frac{x^{r} -1}{r x^{r /2}}  \leq \log x \leq  \frac{2(x^{r} -1)}{r (x^{r} +1)} \leq \frac{x^r -1}{r}.
$$ 
\end{itemize}

\hfill \qed

\begin{Rem}
The bounds of $S(A|B)$ in Corollary \ref{final_corollary} are tighter than those in Corollary \ref{cor01}, whenever the conditions (i) $A \leq B$  or (ii) $B \leq A$ are satisfied.
\end{Rem}



\section*{Acknowledgement}
The author would
also like to thank anonymous reviewers for providing valuable comments to improve the manuscript.
The author was partially supported by JSPS KAKENHI Grant Number 24540146.

\end{document}